**A micromirror array with annular partitioning for high-speed random-access axial focusing**

Nathan Tessema Ersumo[1,2], Cem Yalcin[2], Nick Antipa[2], Nicolas Pégard[3], Laura Waller[1,2,4], Daniel Lopez[5], and Rikky Muller[*1,2,4]

[1] University of California, Berkeley - University of California, San Francisco Graduate Program in Bioengineering, USA

[2] Department of Electrical Engineering & Computer Sciences, University of California, Berkeley, CA 94720, USA

[3] Department of Applied Physical Sciences, University of North Carolina at Chapel Hill, Chapel Hill, NC 27514, USA

[4] Chan Zuckerberg Biohub, San Francisco, CA 94158, USA

[5] Physical Measurement Laboratory, National Institute of Standards and Technology, Gaithersburg, MD 20899, USA

* Corresponding author

**Abstract**

Dynamic axial focusing functionality has recently experienced widespread incorporation in microscopy, augmented/virtual reality (AR/VR), adaptive optics, and material processing. However, the limitations of existing varifocal tools continue to beset the performance capabilities and operating overhead of the optical systems that mobilize such functionality. The varifocal tools that are the least burdensome to operate (e.g., liquid crystal, elastomeric or optofluidic lenses) suffer from low ($\approx$ 100 Hz) refresh rates. Conversely, the fastest devices sacrifice either critical capabilities such as their dwelling capacity (ex: acoustic gradient lenses or monolithic micromechanical mirrors) or low operating overhead (ex: deformable mirrors). Here, we present a general-purpose random-access axial focusing device that bridges these previously conflicting features of high speed, dwelling capacity and lightweight drive by employing low-rigidity micromirrors that exploit the robustness of defocusing phase profiles. Geometrically, the device consists of an 8.2 mm diameter array of piston-motion and 48 µm-pitch micromirror pixels that provide $2\pi$ phase shifting for wavelengths shorter than 1 100 nm with 10-90 % settling in 64.8 µs (i.e., 15.44 kHz refresh rate). The pixels are electrically partitioned into 32 rings for a driving scheme that enables phase-wrapped operation with circular symmetry and requires less than 30 V per channel. Optical experiments demonstrated the array's wide focusing range with a measured ability to target 29 distinct resolvable depth planes. Overall, the features of the proposed array offer the potential for compact, straightforward methods of tackling bottlenecked applications, including high-throughput single-cell targeting in neurobiology and the delivery of dense 3D visual information in AR/VR.

**Introduction**

With the increasingly broad reliance on volumetric processing for improved throughput and precision in optical systems, dynamic axial focusing has recently emerged as an essential feature across several disciplines. Accordingly, varifocal tools have become common fixtures in applications including biological microscopy[1], immersive displays[2], ophthalmoscopy[3], and material processing[4]. Most of these applications involve either coherent scanning systems for volumetric recording, processing and manipulation[1,4] or adaptive focus correction systems destined for coherent imaging or human vision[2,3]. In neurobiology, for instance, coherent optical systems for fluorescence imaging or optogenetic neurostimulation are typically mandated to achieve single-cell resolution targeting. With tissue volumes boasting densities of up to $10^5$ neurons/mm$^3$ and thicknesses of up to 1 mm, such targeting requires dynamic access to several depths at speeds that correspond with the millisecond timescales of neural signalling[1,5,6]. Similarly, in augmented and virtual reality (AR/VR), accommodating depth cues for 3D images entails the use of axial focusing tools[2]. While AR/VR systems typically employ broadband and incoherent light sources, high-speed varifocal tools can seize upon the 1 kHz physiological detection rate of the human eye by rapidly cycling through multiple frames and alternating across coherent light sources to introduce both colour and depth information[7,8]. Thus, by exploiting the persistence of vision using such frame partitioning schemes, high-speed varifocal tools can alleviate the burden posed by the delivery of dense 3D visual information in AR/VR.

Currently, the most prevalent approaches to dynamic axial focusing achieve focus tuning by deforming or reorienting optofluidic[9,10], elastomeric[11] or liquid crystal-based[12] lens components. While such technologies offer straightforward actuation mechanisms, their lagging performance capabilities are increasingly apparent relative to accompanying optical components, especially lateral scanning tools that are often used in conjunction with axial focusing for joint 3D scanning capabilities[1]. Specifically, while optofluidic and elastomeric lenses remain well below the 1 kHz speed threshold needed to achieve submillisecond response times even under optimized conditions[13], state-of-the art lateral scanning tools such as galvanometers routinely achieve refresh rates of tens of kHz[14]. Liquid crystal lenses also suffer from similar speed bottlenecks with the added constraint of having polarization-dependent functionality[15]. In a telling illustration of these stark performance mismatches, recent efforts have even resorted to converting galvanometer-based lateral steering into axial focusing[16].

A general strategy for speeding up axial focusing has been to employ rapidly oscillating systems to continuously sweep across a range of depths. One such approach is the tuneable acoustic gradient index of refraction (TAG) lens, which produces a radial pattern of standing acoustic waves to a fluid chamber to create continuous changes in the refractive index that can achieve focus sweeping at rates on the order of 100 kHz[17]. A second approach takes cue from galvanometric scanners by employing reflective mechanical structures and adapting them to axial scanning by trading tilting resonance modes for ones that produce radial curvature[18,19]. However, the oscillatory behaviour that enables such speeds has also proven to be restrictive as it precludes the capacity for dwelling, which is crucial to applications that require short switching times followed by longer hold durations at specific depths. For example, signal-to-noise ratio (SNR) considerations in some imaging systems impose a minimum bound on the sensor pixel dwell time that often requires several oscillation periods with such continuously scanning tools[20]. Moreover, some optical systems may rely on kinetics that cannot trade optical power linearly against exposure time[21]. Unlike TAG lenses, microelectromechanical systems (MEMS) are opportunely positioned to circumvent this constraint because high mechanical resonance frequencies translate to rapid settling times under DC actuation with optimized drive waveforms[22] or proper damping conditions[23]. Additionally, small and low-mass electrostatically actuated MEMS devices are considerably less susceptible to gravity-induced optical aberrations and less sensitive to mechanical vibrations compared to fluidic or elastomeric systems[24].

Nevertheless, some remaining challenges must be addressed to adapt axial focusing MEMS tools for dwelling-capable operation. Monolithic mirror plates designed to be operated at resonance cannot achieve meaningful actuation under DC actuation without excessively high voltage drives on the order of 100 V or more[18,19]. Furthermore, such structures can often only be actuated in one direction, typically only producing concave curvatures that restrict the resulting dioptric powers to positive ranges[18,19]. One solution to these challenges is to partition the active MEMS array in an annular fashion into independently addressable rings, as illustrated in Fig. 1. This strategy, which has seen increased consideration and adoption in recent years[25,26], reduces the required displacement range and therefore also decreases the driving range by exploiting phase-wrapping capabilities, as each actuated element would only need to produce a total phase shift of 2π. Under such schemes, the dioptric power range is no longer limited by mechanical compliance bounds and driving limitations, but rather the gradual drop in efficiency that comes from applying discrete phase profiles of an increasing gradient[27] as the target depth moves further away from the default focal plane set by the accompanying offset lens (Fig. 1b). Annular partitioning schemes can also tackle recurring issues in wavefront shaping. Namely, the expanded level of control that discrete independent rings provide can eliminate radial aberrations from imperfect phase profiles produced by axial focusing tools[28] as well as spherical aberrations from other components in the optical system[29], both of which would require complex optics for tailored corrections. In addition, annular partitioning can accommodate the requirements for simultaneous targeting of multiple depths by allotting subsets of rings to different target depths[30].

Despite such benefits, annular geometries suffer from drawbacks that have led some to favour alternative partitioning schemes. Specifically, annular concentric structures vary significantly in size, leading to nonuniform actuation behaviour that complicates driving and settling schemes. Moreover, size, shape and suspension schemes across such structures can introduce varying levels of torsional instability and residual stress mismatches resulting in buckling or curling[31]. Given these considerations, a two-dimensional periodic array of identical segmented micromirrors or of identical mechanically intercoupled deformable mirrors, numerous versions of which have been developed over the years[32,33], could offer a better partitioning scheme. Admittedly, periodic geometries increase the proportion of non-active areas across the illuminated aperture region, thereby introducing static diffraction patterns and decreasing the active diffraction efficiency set by the square of the fill factor[34]. However, such impacts can be mitigated by adopting any of the several tactics that are routinely employed today with pixelated spatial light modulators, including fill factor maximization, amplitude masking and spatial filtering[35,36].

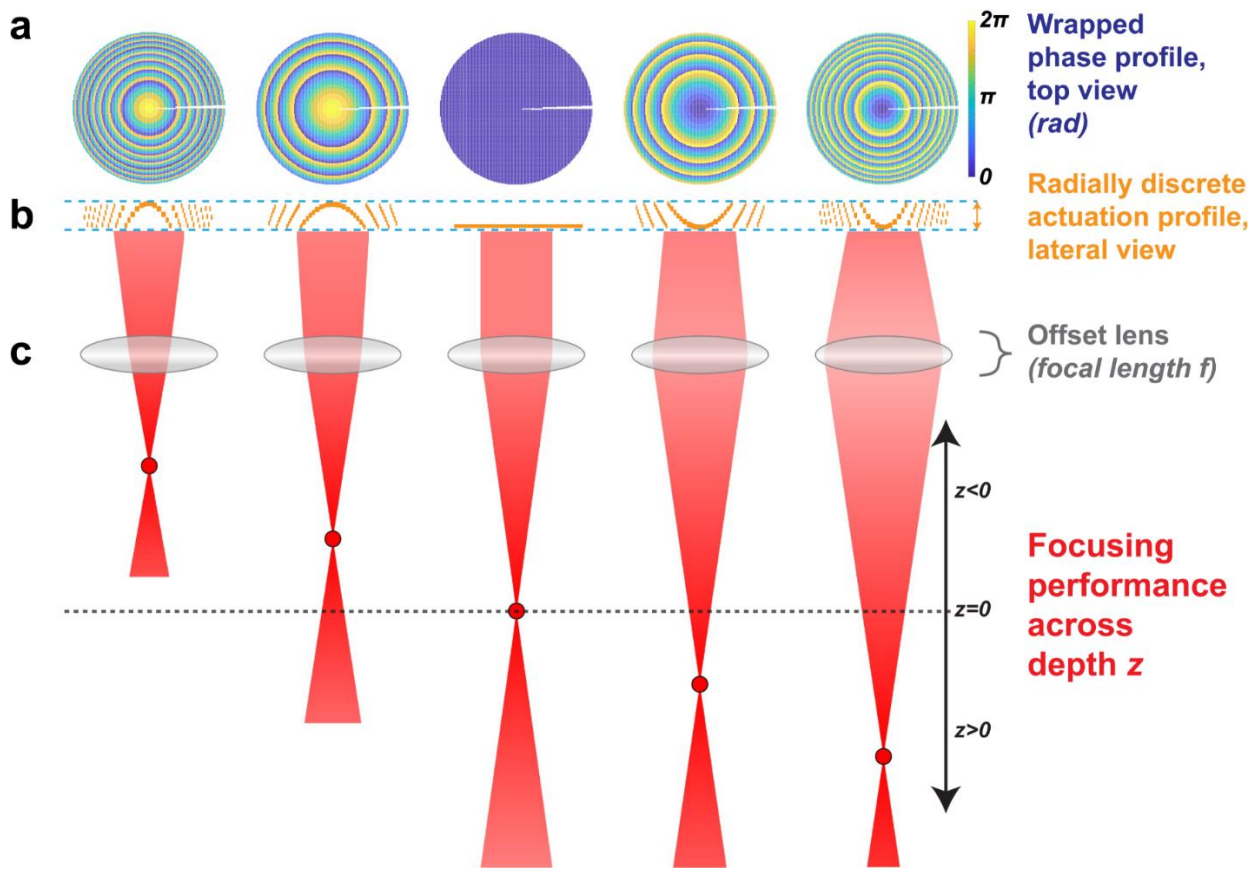

**Fig. 1**. *Schematic diagram illustrating the principle of operation of a radially partitioned varifocal micromirror array. By independently actuating addressable rings, phase-wrapped concave and convex*

*phase profiles may be produced to dynamically focus light to different depths along the optical z axis. An offset lens of a given focal length f is used in conjunction with the array to focus at a default position z=0 when the phase is kept uniform across all the rings of the array.*

In summary, the fast response times and uniform actuation behaviour of small unit structures make pixel partitioning preferable to annular partitioning[32]. However, a survey of currently available MEMS arrays reveals that existing array-based solutions are not ideally suited for adaptation to nimble and general-purpose axial focusing. The broadly used digital micromirror devices (DMDs), for instance, which offer binary amplitude modulation, have been employed to target multiple depths via the generation of Fresnel zone plates, but the generation of foci at symmetric orders and efficiencies on the order of 1 % make such tools impractical for axial focusing[37]. Deformable mirror arrays, on the other hand, are subject to inter-actuator coupling, which impedes radial phase wrapping, and utilize highly rigid suspension schemes that raise voltage drive requirements to hundreds of volts across hundreds of actuation channels, creating substantial operating overhead[38,39].

Hence, the need for a high-speed axial focusing tool with reasonably light operating overhead and features for general-purpose use (including independence from polarization, operability across a wide wavelength range and dwelling capacity) remains unmet. Here, we demonstrate a micromirror-based system that satisfies these requirements by striking a balance between annular partitioning for discrete radial phase control and 2D periodic micromirror tiling for uniform and high-speed actuation behaviour. A circular micromirror array forms the active area of the focusing tool, and simple voltage-driven parallel-plate electrostatic actuation produces the piston motion required for phase shifting across the array's 23 852 micromirror pixels[40,41]. Importantly, the micromirrors were electrically wired into 32 independently addressable annular rings and the micromirror suspension rigidity was optimized for a low-voltage drive (< 30 V). Compact integration to a 32-channel off-the-shelf digital to analogue converter (DAC) therefore allows us to achieve full focusing operation with a straightforward and uniform driving scheme. While the higher sensitivity of the suspension scheme increases the susceptibility to process variations and results in some deviation in actuation behaviour across pixels, the primitive nature of radially varying discrete-step phase profiles allows us to benefit from the averaging effect of having up to hundreds of pixels in a given ring. In addition, the pixel structure and tiling scheme are designed to ensure a systematic wiring process that maintains micromirror planarity under conformal deposition constraints and to provide mechanical stops that prevent electrode contact. Last, mirror stroke is designed to allow for $2\pi$ phase shifting across wavelengths of up to 1 100 nm for a spectral range that encompasses ultraviolet, visible and part of the near-infrared regions.

**Results**

*Pixel-level fabrication*

MEMSCAP's PolyMUMPs and MUMPs-PLUS platforms were used to produce the focusing array, with custom postprocessing performed for reflective layer deposition. The fabrication steps, micromirror structure and tiling geometry are shown in Fig. 2. A unit pixel is a 48 µm-pitch electrically grounded micromirror plate suspended with two clamped-guided beams over a fixed driving electrode. This micromirror geometry was patterned from three polysilicon layers and one gold metal layer for reflectivity, with electrical routing and fixed driving electrodes confined to Polysilicon 0. As part of the semi-custom modifications, the Polysilicon 1 layer, which forms the body of the suspension beams, was thinned down from the 2 µm standard for the process to 0.5 µm to reduce the spring stiffness and lower the voltage drive requirements. To reinforce the body of the micromirror plate and prevent curling due to residual stress mismatches, the suspended mirror bodies were patterned from a double stack of Polysilicon layers 1 and 2. As an accompanying step in stress mitigation, custom evaporation and lift-off postprocessing were performed to reduce the thickness of the reflective gold layer from the 500 nm process standard to 250 nm.

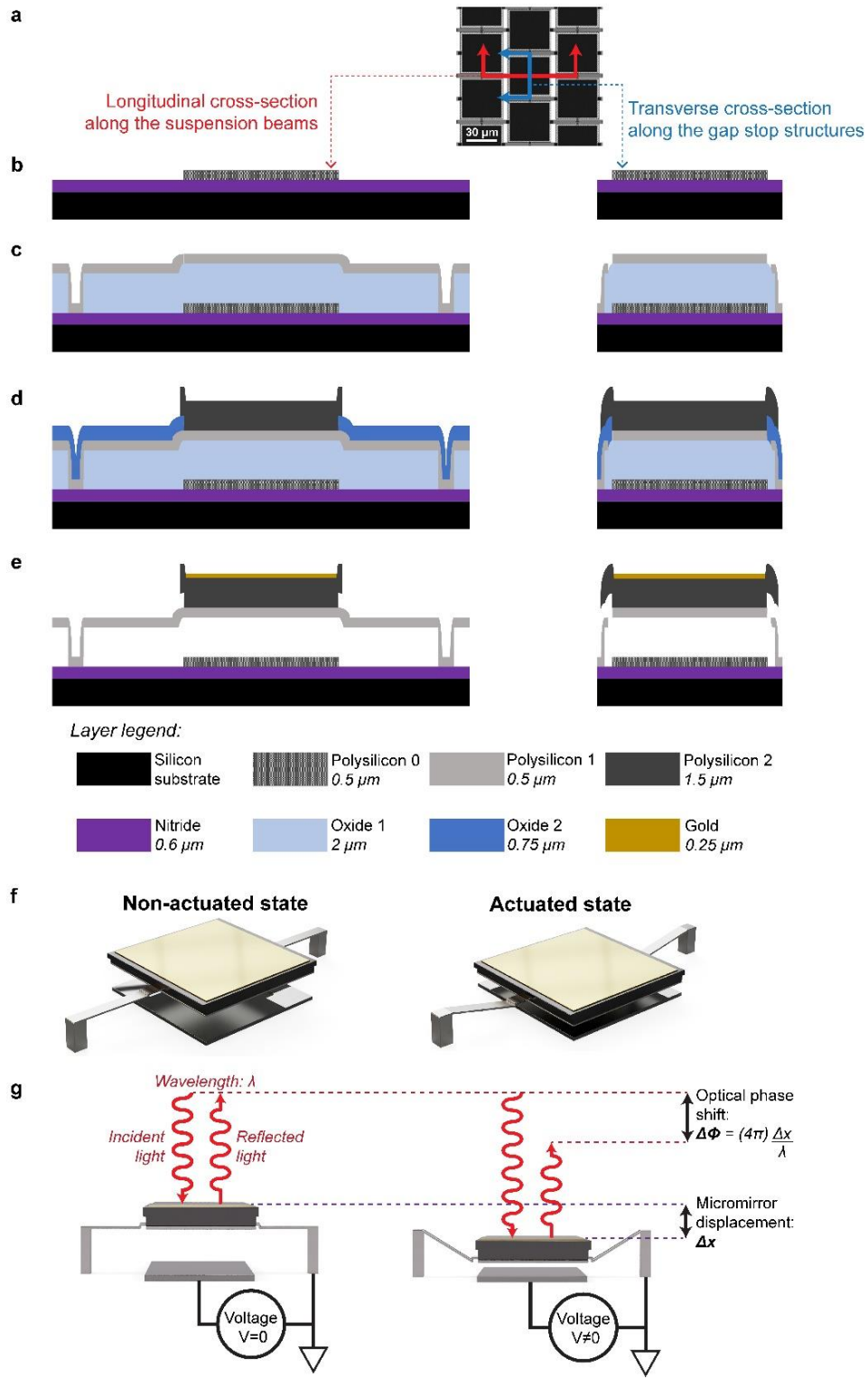

*Fig. 2.* Fabrication and actuation principle of unit micromirror pixels. (**a**) Top-view of the orthogonally staggered pixel tiling geometry with the locations of the fabrication cross-sections in b-e denoted by coloured lines. (**b-e**) Fabrication cross-sections after the deposition and patterning of the following layers: (**b**) Polysilicon 0, (**c**) Polysilicon 1, (**d**) Polysilicon 2, and (**e**) gold. (**f-e**) Isometric and sagittal views of the micromirror renderings at rest and under actuation with an exaggerated scale along the direction of displacement.

The conformal nature of the deposition steps in the fabrication process introduces a top-to-bottom planarity constraint that prevents the patterning of suspension beams and anchors underneath the active region of each mirror plate. Accordingly, a gold metal surface area of 40 µm x 40 µm was allocated for each pixel (Fig. 2e-f), and an orthogonal array format with a staggered tiling scheme was chosen to accommodate non-overlapping suspension beams that extend into adjacent pixels (Fig. 2a). In addition to extending the length of the suspension beams for further stiffness reduction, this tiling scheme prevents electrode contact in the event of electrostatic pull-in or pixel failure. This capability is achieved by having suspension beam anchors serve as gap stops for overhanging juts that are strategically placed to protrude from the Polysilicon 2 layer of adjacent mirror bodies (Fig. 2e). Thus, while the 2 µm size of the electrode gap triggers pull-in past a mirror displacement of 667 nm under electrostatic voltage drive, the mechanical stops cap the maximum displacement at 750 nm, at which point a reduction in the applied voltage below the pull-out threshold restores the mirror back to a regular operating regime. The overhanging juts were sized to be large enough to accommodate the requisite mask misalignment tolerance for the fabrication process but small enough to avoid contact with neighbouring juts such that mirror bodies were physically isolated from each other at all times. Altogether, our pixel-level micromirror structure and complementing tiling scheme efficiently exploited the space and material made available by the fabrication process to produce a robust actuation scheme that abides by planarity and feature size constraints. The relationship between this electrostatic actuation scheme and optical phase shifting is illustrated in Fig. 2g: the piston-motion actuation of a given mirror increases the optical path of locally incident light, adding twice the actuation displacement to the travel distance.

The top views of the fabricated arrays in Fig. 3 further illustrate how the chosen pixel and tiling geometries also incorporate an efficient architecture for pixel wiring with a minimal footprint. Traces connecting rings to bond pads were placed together within a dedicated 7.2° radial slice to minimize routing placement overhead and keep area usage consistently at 2 % regardless of the chosen size and pixel count of the circular array (Fig. 3a,b). Importantly, wiring between adjacent pixels belonging to the same ring is ensured via 8 potential pixel-level connection points whose placement/omission does not impact mirror planarity, as shown in Fig. 3c. This connection scheme also allows for an automated placement process during layout once the exact ring partitioning geometry is chosen. Last, as part of our structural inspection of the fabricated arrays, topography measurements were performed using atomic force microscopy (AFM) to evaluate the quality of the custom-deposited reflective gold layer. No planarity issues were noted, and the height variation across the 50 nm x 50 nm unit regions of the gold surface had a standard deviation of 12.09 nm.

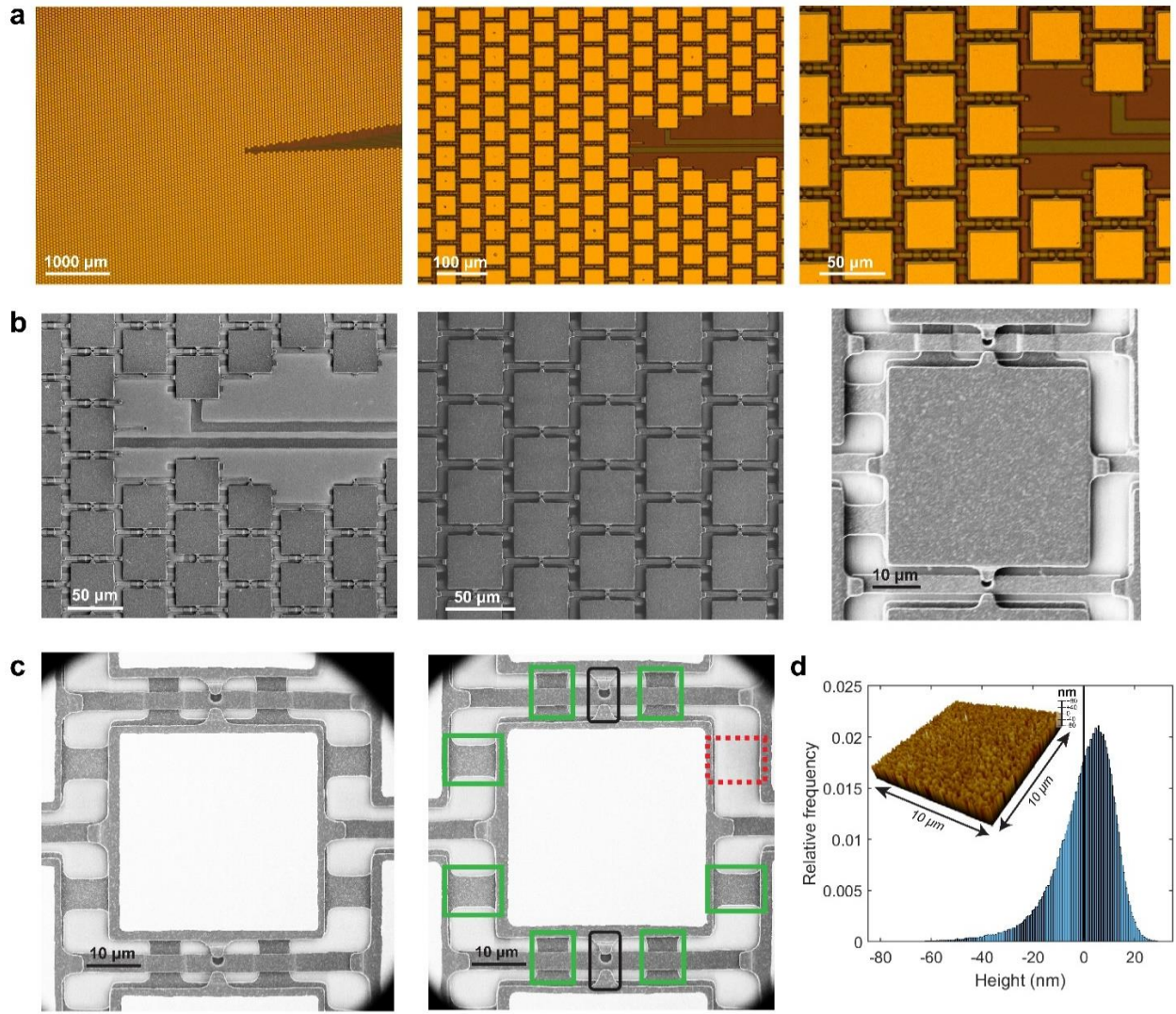

*Fig. 3*. Fabrication images of the produced array. (*a*) Optical microscopy images of the array centre at various levels of magnification. (*b*) Scanning electron microscopy (SEM) images of the array prior to metallization postprocessing at various levels of magnification. (*c*) SEM images of two different pixels after metallization. Electrical wiring across pixels for grouped drive is achieved via 8 potential connections between fixed bottom electrodes at the Polysilicon 0 layer, which can be added (green boxes, right image) or removed (red box, right image) in accordance with the partitioning geometry. For each pixel, two Polysilicon 2 juts (black boxes, right image) hanging over the anchors of adjacent suspension beams serve as gap stop structures preventing electrode contact from electrostatic pull-in

*during operation. (**d**) Topography histogram of the deposited reflective gold layer as measured by atomic force microscopy (AFM). The heights are measured per 50 nm x 50 nm region; histogram bin sizes are 0.5 nm. The top-left inset is a sample 2-dimensional topography plot across an area of 10 μm x 10 μm.*

*Array-level fabrication*

The circular array was chosen to have a 32-ring partitioning geometry and an aperture diameter of 8.2 mm (for a total count of 23 852 pixels) based on an iterative optimization process using a previously described optical simulation framework that can assess the axial focusing range with respect to the axial resolution[42]. A 32-ring addressing scheme was chosen to maximize the compatibility with a 32-channel commercial DAC system that was selected for array driving. As shown in Fig. 4a, ring-level track widths were gradually reduced with increasing ring radius to account for the sharper radial gradients produced at the edges of the array during axial focusing. Impedance measurements were also performed for each ring to assess driver requirements and inspect for shorting. In accordance with the parallel plate structure of the micromirrors, the measured impedances were purely capacitive, with the capacitance of each ring closely following the pixel count. The capacitance was generally found to increase with ring radius, and therefore, it also increased with ring area (Fig. 4b). Capacitance drops were noted between the consecutive rings where the track width was reduced. A linear regression of the measured capacitance to the pixel count had an $R^2$ value of 0.98, further illustrating the correlation between the two properties. From the regression, the mean pixel capacitance was evaluated to be 0.22 pF, whereas the mean parasitic capacitance due to traces and bond pads (seen in Fig. 4c) was 26.9 pF/ring. The measured pixel impedance is largely due to the mutual capacitance that exists across the nitride layer between the driving electrode in the Polysilicon 0 layer and the underlying single-crystal silicon, which was grounded via nitride breach structures demarcated in Fig. 4c. A Polysilicon 0 layer wall surrounding the entire array was also deposited to maintain uniform actuation behaviour by shielding micromirrors from any potential residual stress mismatches across the bottom-most layers of the fabrication process[31] as well as to provide a ground connection to all the micromirror bodies. A post-assembly image of a single 1 cm x 1 cm array chip is shown in Fig. 4d.

*Pixel-level and ring-level performance*

Following the assembly of the array chips, pixel-level functionality was evaluated using digital holographic microscopy[43]. The variation in the resting height of pixels across the array was found to have a standard deviation of 13.83 nm, indicating that the impact of beam buckling is minimal relative to the target displacement range of 550 nm. The steady state micromirror displacement was then measured as a function of the applied voltage. To compactly quantify the actuation behaviour of each pixel, the results were fit to a generalized form of the analytical solution for parallel plate capacitive transduction:

$$V = \sqrt{ax(b - \Delta x)^2}$$

This fit reduces all relevant geometric and material parameter values of each pixel down to two parameters *a* and *b* that accurately capture the differences in actuation behaviour due to regional process variations, as evidenced by the fact that all the $R^2$ values exceed 0.99. Based on the average actuation behaviour, the mean applied voltage for a displacement of 550 nm, i.e., the 2π phase shift at a wavelength of 1 100 nm, was 29.65 V, while pull-in was found to occur at a mean voltage of 30.35 V. The parameter *b*, which corresponds to the effective electrode gap distance, was also assessed to be 2.16 µm, an ≈ 8 % deviation from the nominal process value of 2 µm. The steady state actuation measurements for individual pixels are shown in Fig. 5a along with the mean behaviour and 95 % confidence intervals across the full displacement range.

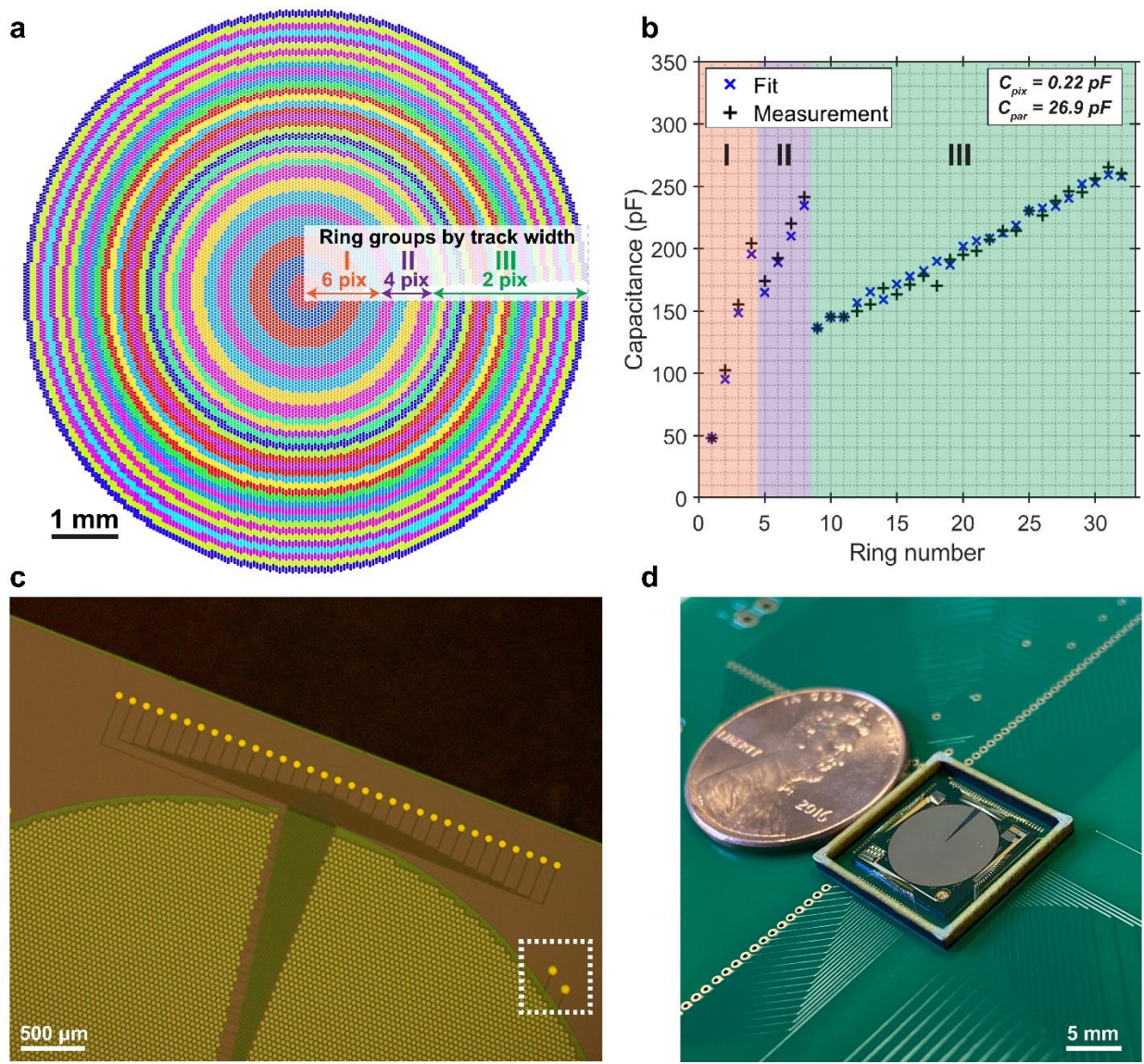

*Fig. 4.* The selected wiring scheme determines the driving requirements for the array. *(a)* Fabricated arrays were partitioned into 32 rings with track widths that decrease with increasing radius to account for the radial slope of the applied spherical phase profiles. *(b)* Measured capacitance measurements of each ring. *(c)* Microscopy image showing the employed wiring scheme. Nitride breach structures (demarcated here with a white dashed box) were also used to ground the substrate. *(d)* Photograph of the mounted and assembled array chip next to a US penny.

The dynamic pixel behaviour was subsequently characterized by measuring settling responses to various voltage steps under a stroboscopic setup, as shown in Fig. 5b. All voltage steps were set to have an amplitude of 10 V, and the nonlinear nature of the actuation behaviour was exploited to achieve varying magnitudes of displacement by modifying the starting voltage offset from 0 V to 5 V and 20 V. The obtained measurements reveal an overdamped response with a settling duration that remains fairly consistent across changing magnitudes and directions in the voltage step. Under a 2 % settling time metric, the mean response time was measured to be 114 µs for a refresh rate of 8.75 kHz. Under a 10-90 % settling metric for suitable comparison against alternative approaches to axial focusing, the mean response time was measured to be 64.8 µs for a refresh rate of 15.44 kHz, which is roughly two orders of magnitude faster than the current commercial optofluidic and liquid crystal-based varifocal systems[9,15]. Overall, these response measurements demonstrate operating speeds that match the 10 kHz benchmark achieved by galvanometer mirrors and that could be raised even further under optimized damping or drive shape conditions.

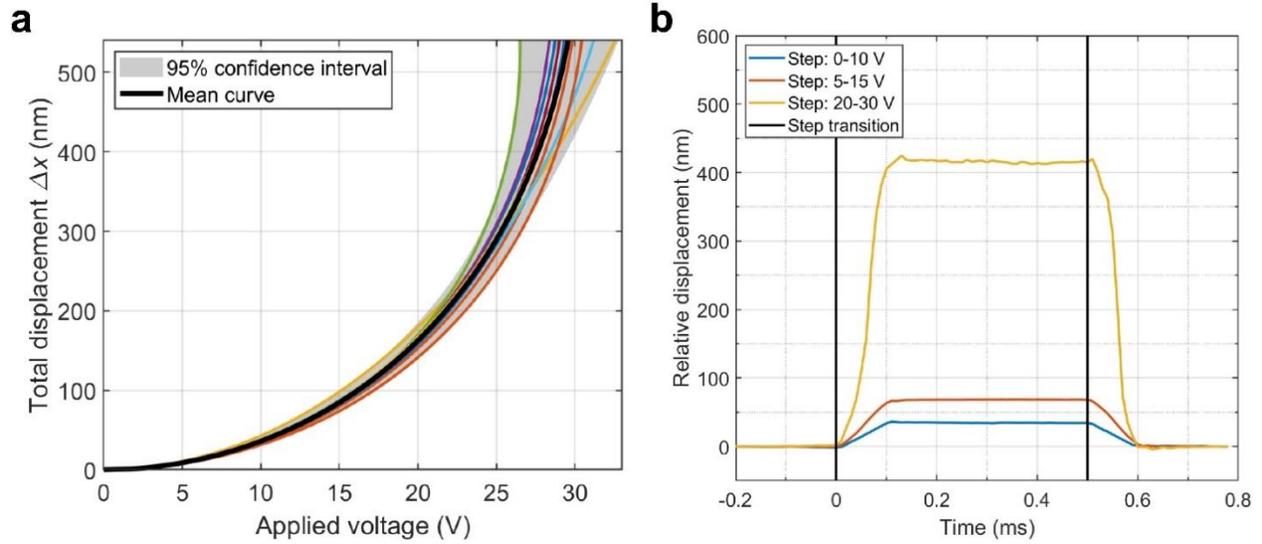

*Fig. 5*. Pixel-level steady state and dynamic transduction characterization results obtained via digital holographic microscopy. (*a*) Measured steady state displacement vs applied voltage. The coloured curves correspond to individual pixel measurements from various pixels across arrays. (*b*) Mean settling behaviour of pixels measured stroboscopically by applying 1 kHz square wave voltage signals with a 10 V step and varying the offsets for increasing ranges of displacement.

The ring-level operation was also visualized across a large field of view (FOV) under digital holographic microscopy, as shown in Fig. 6. While the increased FOV could not sufficiently resolve the micromirror features to accurately reconstruct and quantify the pixel-level phase, the ring-level actuation could be qualitatively evaluated for a single ring subjected to stepwise increases in the applied voltage (Fig. 6a-f) as well as for several rings actuated with the same voltage (Fig. 6 g-h). Altogether, these phase reconstruction images serve to confirm that the employed wiring scheme results in coordinated and uniform co-actuation between pixels belonging to the same ring.

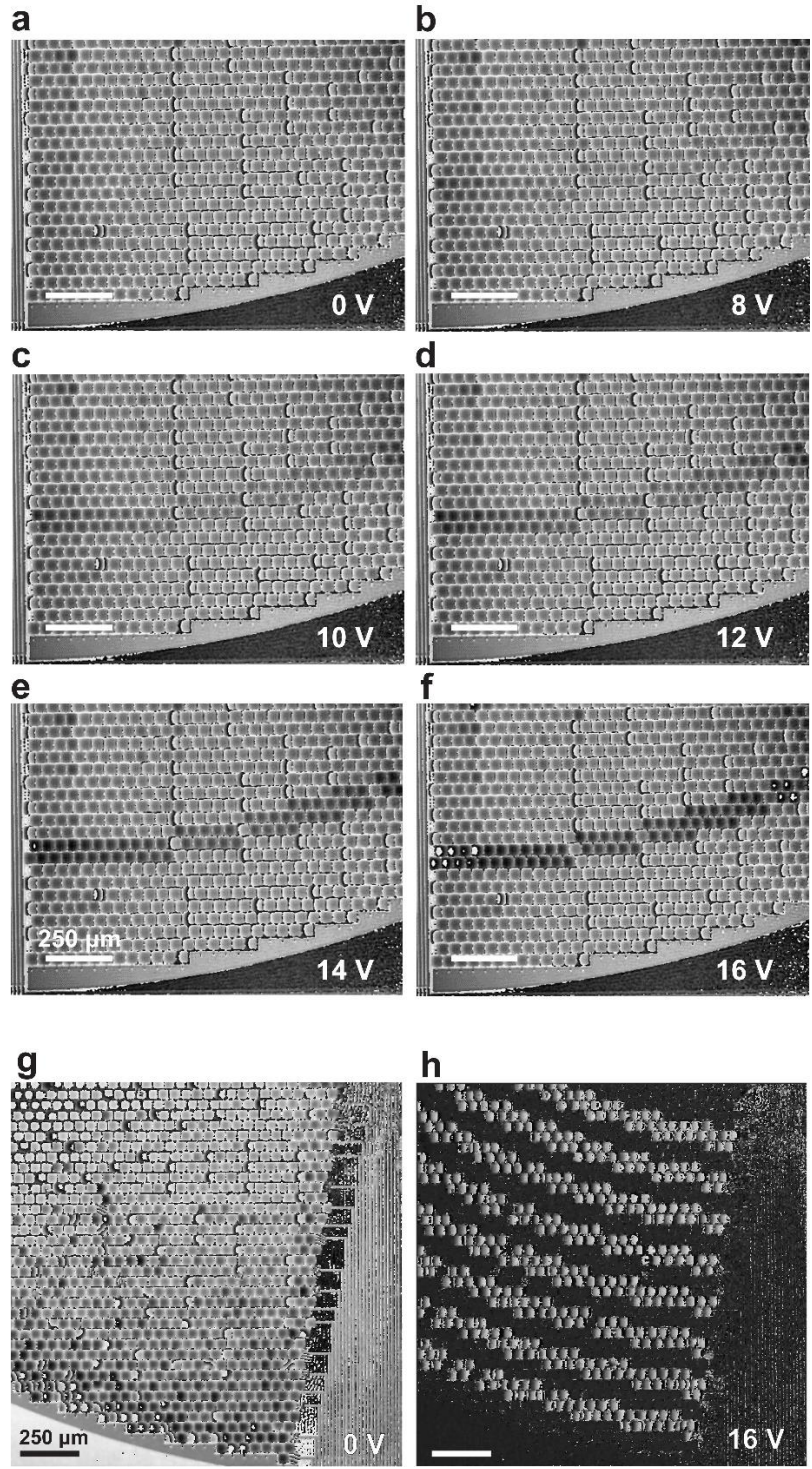

*Fig. 6. Phase reconstruction images obtained via digital holographic microscopy demonstrating ring-level actuation of the array (all scale bars are 250 μm). (a-f) A single ring (#28 of 32 from the centre) was actuated in increasing steps with applied voltages of 0 V, 8 V, 10 V, 12 V, 14 V, and 16 V. (g-h)*

*Alternating rings (#17, #19, #21, #23, #25, #27, #29, and #31 from the centre) were concurrently actuated with an applied voltage of 16 V. (**g**) Phase reconstruction before actuation. (**h**) Phase difference after actuation.*

*Array-scale axial focusing performance*

Once pixel-level and ring-level functionalities were verified, array-scale axial focusing performance was evaluated using the test setup illustrated in Fig. 7. The optical setup consisted of a collimated illumination subsystem involving two laser sources (532 nm and 980 nm wavelengths) as well as a *2f* optical configuration around an offset lens L2 (100 mm focal length) with the micromirror array at the front focal plane. A CMOS camera mounted on an automated z-stage and centred at the rear focal plane of L2 was used to acquire z-stacks for each phase profile applied using the array. To generate these phase profiles, the array was driven using 32 DACs with a 14-bit level of precision that can accommodate the sensitive higher-voltage region of the nonlinear micromirror actuation curve. The mapping between the desired phase and applied voltage was adjusted to the wavelength of the selected laser source such that a total drive of $\approx$ 29 V was required at 980 nm and $\approx$ 24.3 V was required at 532 nm. From the acquired z-stacks, lateral and axial projections of peak intensity values were used to quantify performance metrics including axial spot size, focusing range, lateral spot size, and deviations from the optical axis and target depth planes. Two folding mirrors M1 and M2 were used to multiplex the test setup operation across four configurations, with M1 selecting between the two laser sources. Positioned to fold into place right in front of the array, M2 generated a single passive focus spot (at the rear focal plane z=0), which was used as a reference to assess the insertion loss of the micromirror array.

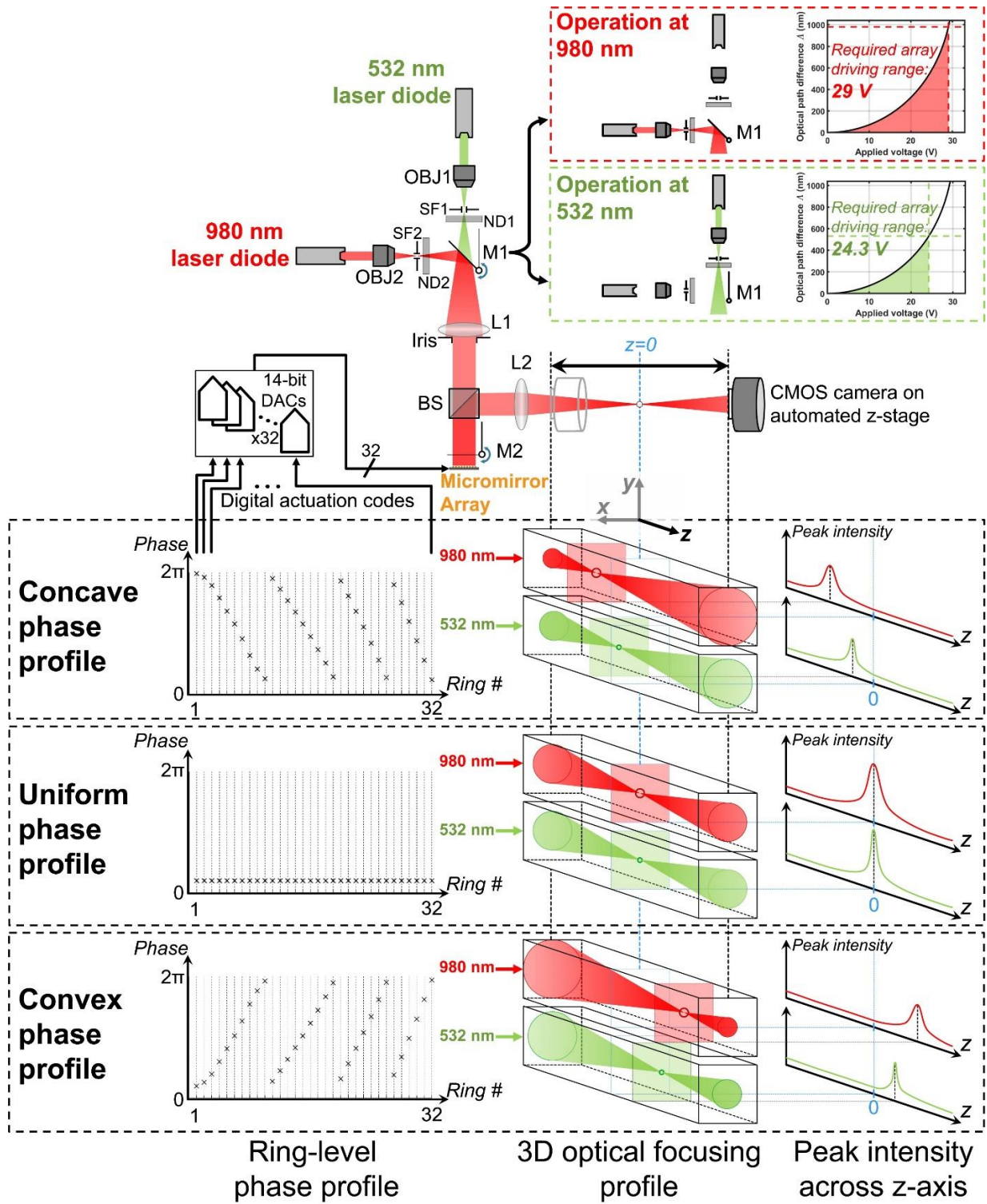

**Fig. 7.** Schematic diagram of the optical test setup, performed measurements, and drive electronics. For two distinct laser source wavelengths (532 nm and 980 nm), multiple phase masks corresponding to

*different target depths along the optical z axis were applied by driving the rings of the micromirror array with 32 DACs. Mirror M1 was employed to switch between laser sources. Mirror M2 was employed to alternate the reflection between the micromirror array and plain mirror to assess the focusing efficiency η. A CMOS camera mounted on an automated z-stage was employed to acquire z-stack images across the full focusing range. For a given convex or concave phase mask, operation at 980 nm resulted in a greater focusing power and a larger spot size compared to operation at 532 nm. (OBJ: objective, SF: spatial filter, ND: neutral density filter, M: mirror, L: lens, and BS: beamsplitter).*

Despite the testing specificity imposed by the choice of the laser wavelengths and a particular focal length $f$ for L2, system- and magnification-agnostic metrics may be extrapolated by exploiting the colinear scaling that governs the focusing range and spot size. Focusing is determined by the following relationship:

$$d_o \cdot d_i = f^2$$

where $d_o$ is the distance from the micromirror at the rear focal plane to the focus spot in the object plane, and $d_i$ is the distance from the rear focal plane to the focus spot in the image plane. The axial spot size is proportional to:

$$n\frac{\lambda}{NA^2} \propto \frac{\lambda \cdot f^2}{n \cdot D^2}$$

where λ corresponds to the laser wavelength, $n$ corresponds to the medium's refractive index, NA corresponds to the system's numerical aperture, and $D$ corresponds to the aperture diameter set by the size of the micromirror array ($NA \approx 0.5\ nD/f$ under the low NA regime that is being considered). Accordingly, given the mutual scaling with $f^2$, the ratio of the axial spot size to the axial focusing range is intrinsic to the micromirror array and serves to measure the number of distinct, resolvable depth planes that the array can produce. Moreover, the focusing range is bounded by the extent of 2π-wrapping present in the phase profile produced by the array, which in turn scales with the wavelength λ, as evidenced by the following relationship determining the phase shift ΔΦ required at a location (*x,y*) on the array:

$$\Delta\Phi(x,y) = 2\pi \frac{d_0 - d_0\sqrt{1 - \frac{(x^2 + y^2)}{d_0^2}}}{\lambda}$$

Thus, in addition to being agnostic to focal length *f*, this range-to-spot size metric remains conserved across wavelengths assuming minimal impact from non-idealities in the array and lens L2. For this metric, the spot size is quantified as the full width at half maximum (FWHM), while the axial focusing range is quantified as the range across which the peak intensity of the desired focus spot exceeds that of the undesired diffraction effects.

The simulated and measured axial focusing performance at both 532 nm and 980 nm wavelengths are presented in Fig. 8a,b. The decreased efficiency that accompanies the higher phase gradients produced when targeting depths further away from the rear focal plane[27] can be observed in the simulation results and is reflected in the experimental data. This indicates that the efficiency is primarily dependent on the discrete nature of the ring-level phase steps and not the pixel-level variation in the actuation behaviour. While the simulation places zeroth-order diffraction efficiency under a uniform phase profile at 52 %, consistent with the principle that it should theoretically be equal to the square of the fill factor[34], the measured efficiencies at 532 nm and 980 nm were 12.3 % and 24.3 %, respectively. These additional losses can be attributed to the reflectivity of the thinned gold layer, as previous characterizations of gold films of similar thickness are in agreement with the fold changes in the efficiency seen between the measurement and simulation at both wavelengths[44]. A three-dimensional visualization of the focusing performance is also provided with the lateral projections of acquired z-stacks in Fig. 8c,d. The appearance of faint spots at locations that are bilaterally symmetric to the target depth planes with respect to the rear focal plane is most likely the result of a small subset of hypersensitive pixels in each ring that effectively behave much like zone plates under binary operation[37].

The quantifications of the axial and lateral positions of the generated spots (Fig. 8 e, f) demonstrate the absence of significant lateral deviations from the optical axis as well as good agreement between the targeted and obtained depths. The appearance of some slight deviations in the axial position around the most negative target depth values also illustrates how calibration using such measurements provides a straightforward opportunity for the micromirror array to identify and tackle any system-specific rotationally symmetric aberration via the adjustment of a lookup table for ring-level actuation codes. Finally, with the axial spot size across the focusing range measured to be 929 µm ± 195 µm (mean ± standard deviation) and the focusing range evaluated to be 27.5 mm at a 532 nm wavelength, the axial range-to-spot size ratio is calculated to be ≈ 29.6. While this represents a drop from the simulated ratio value of 37.6 resulting from a slight degradation in the axial spot size due to pixel-level variations in actuation behaviour, the array's demonstrated ability to resolve these many distinct depth planes comfortably meets the requirements across several applications in fields including biological microscopy[45] and material processing[4].

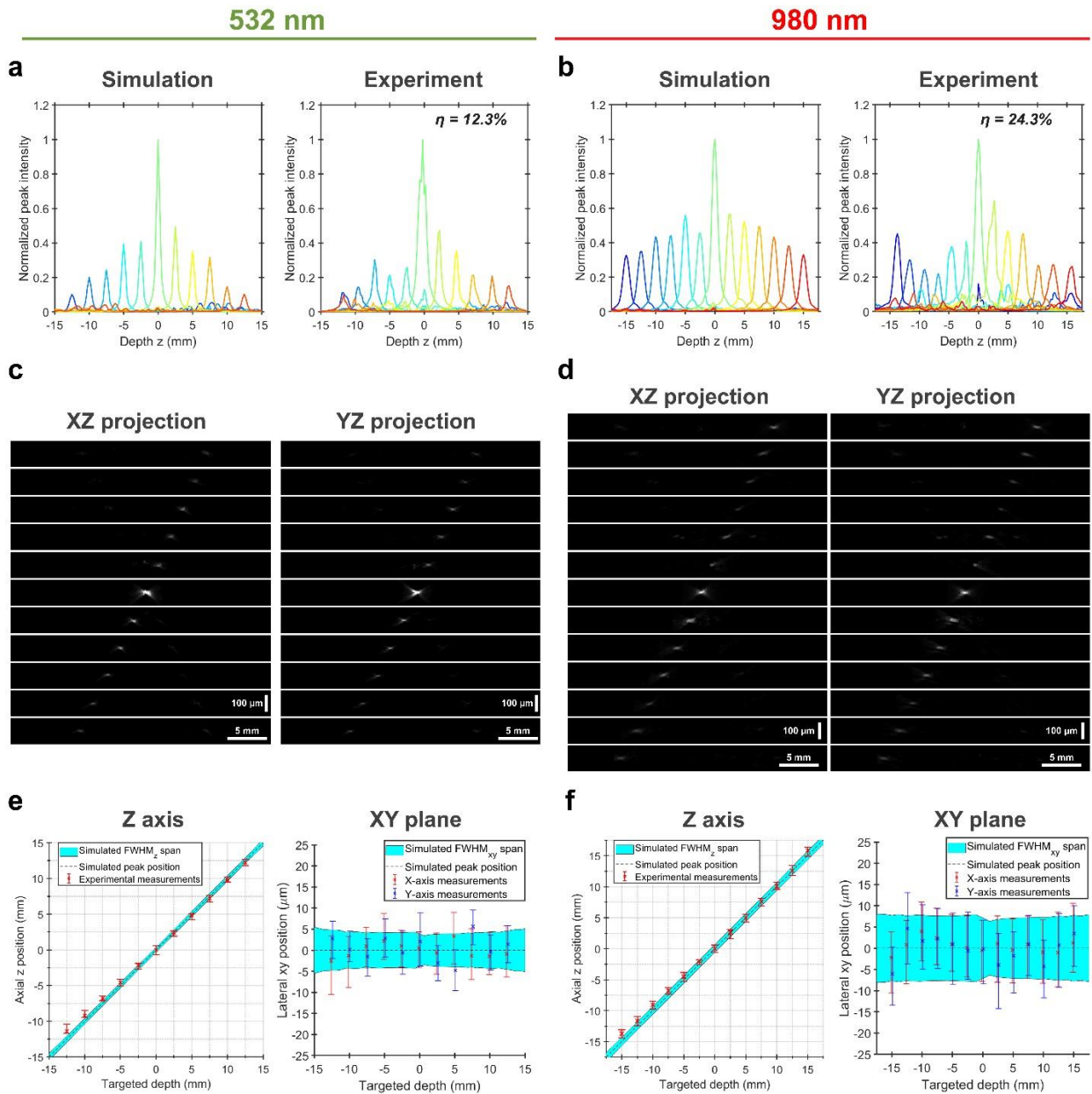

***Fig. 8.*** *Optical focusing performance results. **(a-b)** Simulated and experimental results of the peak intensity values projected along the optical axis for 532 nm **(a)** and 980 nm **(b)** sources. Each coloured line corresponds to a distinct applied phase mask, and the efficiency η is included as an inset. **(c-d)** Lateral projections of the acquired Z-stacks along the XZ plane and YZ plane for all applied phase masks with 532 nm **(c)** and 980 nm **(d)** sources. The scales for the lateral dimensions (X and Z) differ from the Z-axis scales as indicated by the provided scale bars. **(e-f)** Spot size and position measurements along the*

*X, Y and Z dimensions compared against the simulated values for 532 nm (**e**) and 980 nm (**f**) sources. The spans of the measurement bars and simulation envelopes correspond to the spot size values measured as full width at half maximum. X-shaped measurement markers correspond to the peak intensity position along z for left side plots and the peak intensity distance from optical axis for right side plots.*

**Discussion**

Whether applied under an adaptive optics framework for defocus correction or in a 3D translation context for depth targeting, axial focusing constitutes a fundamental and ubiquitous mode of optical manipulation. Fittingly, the micromirror array presented in this work achieves axial focusing capabilities suitable for general-purpose use by exploiting the robustness of this fundamental phase mode to alleviate the driving burden. By adopting an architecture that reserves phase accuracy for the outer regions of the active area, allows for radial phase-wrapping, and relaxes uniformity in favour of sensitivity, the described array reached a refresh rate of $\approx$ 15 kHz across wavelengths of up to 1 100 nm, with only 32 addressing channels and < 30 V of required drive. Furthermore, this performance was achieved without the hindrance of constraints such as polarization dependence, continuous sweeping and non-ideal radial phase curvatures. Unlike optofluidic and elastomeric lenses that rely on refraction for focus tuning, the developed array employs annular phase shifting. While this mechanism of optical modulation does result in wavelength-dependent operation, we note that varifocal performance capabilities are often destined for coherent applications, including fluorescence-based imaging, scanning optogenetic manipulation, and laser micromachining[1,4]. Since the array can accommodate a wide range of wavelengths, its high refresh rate can also be employed to achieve temporally multiplexed operation across several wavelengths for polychromatic applications with temporal integration scales of less than $\approx$ 10 kHz. For instance, in AR/VR, where varifocal tools are mandated for the prevention of vergence-accommodation conflicts[2], the proposed tool can be especially nimble in exploiting the window across which the human eye integrates kilohertz-speed physiological detections for an effective perception rate of $\approx$ 100 Hz.[7] Specifically, the array could be used to either partition complex 3D images into simpler frames, couple in lateral scanning tools, or multiplex light sources, all without any restriction on the order and duration of the targeted depths[7,8].

In addition, while the overall peak efficiency of the array was measured to be in the range of $\approx$ 10-30 %, a substantial portion of these losses are incurred at the array plane due to the low reflectivity of the gold layer[44], which was thinned down to minimize the likelihood of mirror curling. However, given that no substantial mirror curling was observed, the thickness of this reflective layer may be increased for improved reflectivity[44]. More importantly, since the remaining losses contributing to the measured efficiency are not the result of unwanted scattering across the target zeroth-order region but rather because of the power being diverted to higher diffraction orders outside of the FOV, the optical power incident on the array may be increased to produce brighter images for applications that are not power-limited. Off-target artefacts within the zeroth-order region will remain minimal during operation because the reported axial operating range is by definition measured as the tuning range across which off-target artefacts are not significant and not as the full tuning range set by the physical limits of the array. Moreover, by contextualizing this range with respect to the spot size in the form of a ratio, we demonstrate that the array's wide tuning range does not come at the expense of resolution. For instance, the achieved axial range-to-spot size ratio of $\approx$ 29.6 exceeds the requirements imposed by optical systems involving either neural stimulation or recording, where neuron targets have sizes on the order of 10 µm and scattering under linear one-photon regimes limit the accessible depth to $\approx$ 100 µm.[5] More pertinent to state-of-the-art systems, the axial confinement offered by multiphoton regimes that extend accessible depths to $\approx$ 1 mm can also be exploited to relax the axial resolution and achieve single-cell full-range targeting at speeds that exceed the characteristic $\approx$ 1 kHz benchmark of neural signalling[6]. Overall, the array's performance capabilities can thus potentially eliminate bottlenecks across several applications.

Despite the high speeds associated with MEMS-based approaches, the inertial nature of mechanical moving systems has spurred parallel efforts favouring solid-state mechanisms for optical manipulation[14]. Namely, acousto-optic and electro-optic modulation approaches, which offer refresh rates of up to 1 MHz,[14] have become increasingly popular in recent years with the widespread adoption of tools such as acousto-optic deflectors for lateral scanning[46]. Unfortunately, this benefit of ultrafast responsivity has not translated into viable high-speed axial focusing, as limitations in the control and sensitivity result in performance costs such as drastic insertion loss[47] or high voltage drive requirements[48]. In addition, MEMS structures have the distinct advantage of being ideal dynamic substrates for metamaterials. Functional metasurfaces such as metalenses offer the potential for compact and highly tailored functionality in future generation optical systems. However, their inherently passive nature requires the use of complementing active elements, which typically become sources of performance bottlenecks. With MEMS-based dynamic substrates, such shortcomings related to response times and fabrication compatibility can be circumvented[49,50].

While the strategy of increasing the micromirror sensitivity for reduced voltage drive in this work does lead to more variation in the actuation behaviour across pixels, the reported functional testing results demonstrate that these mismatches do not substantially affect the axial focusing performance. These pixel-level disparities in actuation can be directly attributed to geometric and process-related properties, such as the electrode gap thickness, suspension beam thickness, and residual stress. Since die-level variations of these properties exhibit spatial continuity, the potential for regional drive correction (akin to calibration approaches in full-fledged spatial light modulators)[51] opens up opportunities for expanded phase control using similar low-voltage pixel structures.

Another design tradeoff concession made to maximize the fill factor while abiding by planarity constraints was the use of two suspension beams per pixel instead of three or four. This structural limitation, which introduces an unconstrained torsional degree of freedom, was mitigated with the incorporation of gap stops to prevent electrode contact in the event of tilting and was not found to substantially degrade the lateral spot size along any specific orientation. Similar to pixelated spatial modulator arrays, the presented axial focusing array's reliance on discrete phase steps results in target depth-dependent efficiency[27]. Although this efficiency profile may not be ideal for applications that require uniform intensity across depth, the axial operating range can be restricted to a region across which the efficiency variation is less extensive, as evidenced by the 980 nm focusing performance results (Fig. 8b). While the overall impact of torsional freedom and efficiency on the functional performance of the varifocal array remains limited, the need for concessions such as these can be obviated in future design iterations by expanding the fabrication process. Specifically, a planarization step could decouple the suspension network from the active area, thereby offering increased space for both additional suspension beams and higher fill factors.

Altogether, the axial focusing tool presented in this work constitutes both a versatile product and an attractive platform for expanded or reconfigured optical modulation. For the current array embodiment, the light operating overhead that the 30 V, 32-channel driving scheme represents makes compact on-board driver integration feasible. With a reduced wavelength range requirement, driving ranges can be lowered even further by applying a negative voltage bias to suspended mirror bodies: halving the wavelength range from 1 100 nm to 550 nm would, for instance, reduce the driving voltages to less than 8 V under this approach. Moreover, the systematic nature of the pixel wiring process offers a straightforward pathway from general-purpose to application-specific design that may involve modifications in the aperture size, reconfigurations to the partitioning scheme, or readjustments to the number of addressable channels[42]. Finally, this micromirror platform can potentially accommodate expanded features such as limited dynamic partitioning via relays outside of the active area.

**Materials and methods**

*Design, fabrication and assembly of the array*

The array-scale geometry of the axial focusing tool was designed using a previously described custom simulation framework[42]. Reflective pixel elements were placed in the active region of the computational framework in accordance with the chosen tiling scheme and sorted into elemental rings. The final 32-channel partitioning arrangement and aperture size were chosen based on an iterative process involving regrouping these elemental rings and evaluating their focusing performance. The pixel-level micromirror structure was developed using analytical simulations of parallel-plate capacitive transduction as well as finite element analyses. Following fabrication and oxide release, array chips were mounted and wire-bonded to custom printed circuit boards (PCBs) via epoxy-based attaching and gold ball bonding.

*Micromirror array inspection*

Released, standalone array chips were examined under optical microscopy, scanning electron microscopy, and atomic force microscopy. AFM was performed in noncontact mode to assess the topography of the deposited reflective gold layer. The heights were measured per 50 nm x 50 nm region across several pixels and a total area of 190 $\mu m^2$. The impedance measurements were also performed on released array chips at a frequency of 1 kHz and under a parallel $C_p$-$R_p$ model using an LCR Meter and a micromanipulator probe station for direct access to the bond pads.

*Array driving*

The micromirror array was driven with a 32-channel 14-bit precision DAC. A commercial FPGA programmed with custom firmware was used for digital interfacing between the DAC and software for operation. Digital and analogue power supplies were provided to the DAC using a power supply and a precision source/measure unit. The 32 voltage outputs of the DAC (along with ground) were connected to the micromirror array PCB via ribbon cable.

*Digital holographic microscopy*

Digital holographic microscopy was performed on assembled arrays in ambient air at room temperature for phase and amplitude reconstruction using a 675 nm wavelength laser source under a reflection-based interferometric scheme[43]. While steady state measurements were performed using DAC voltage outputs, dynamic measurements were performed using a stroboscopic unit, which employed a dedicated voltage output (10 V maximum swing with 2 % accuracy) for synchronization to laser pulses. Offsets of 0 V, 5 V and 10 V were applied with a 10 V amplitude to access different regions of the nonlinear micromirror actuation curve and evaluate the step responses across small and large displacement ranges. The stroboscopic unit achieves precise measurements of megahertz-regime mechanical actuation speeds by applying a periodic drive, applying nanosecond-regime laser pulses at specific time offsets of the periodic signal, and integrating the signal from several periods spanning the camera's shutter time for adequate SNR. Specifically, micromirror settling behaviour was reconstructed from stroboscopic acquisition runs employing a 1 kHz square wave driving signal and with an effective sampling rate of 100 kHz.

*Optical testing of the focusing performance*

The functional performance of the axial focusing array was evaluated under the optical test setup illustrated in Fig. 7 in ambient air at room temperature. A camera was mounted on a motorized linear stage for 12-bit z-stack acquisitions of each applied phase. The step size of the linear stage, i.e., the axial precision of the acquisition was set to 200 µm. The FPGA-mediated DAC voltage drive, camera acquisition and stage control were coordinated for automation using custom software. Collimated laser modules were used as the 532 nm wavelength and the 980 nm wavelength laser sources. An achromatic doublet with a focal length of 100 mm was used as the offset lens for axial focusing. To ensure measurement accuracy, four acquisition runs were performed and averaged for each applied phase profile. The efficiency was assessed by integrating the pixel intensities across the zeroth-order spot produced by the micromirror array as well as the focus spot produced by the plain reference mirror. The efficiency was subsequently calculated by taking the ratio between the two integrated intensities and multiplying this ratio with the reflectance of the plain mirror at the wavelength under study. For each wavelength, a geometric mean of z-stacks across all the applied phase profiles was used to subtract out the background, back-reflections from flat optical elements and static reflections from non-active areas of the array chip. The optical axis was determined in the z-stacks by performing linear regressions of the X and Y positions of the peak intensity values from each applied phase profile to their depth position Z. The lateral spot size was obtained at the peak intensity depth plane by estimating the FWHM along the X and Y axes via spline interpolation. Similarly, the axial spot size was determined for each target spot by generating a profile of the peak intensity at each depth and estimating the FWHM from this profile via spline interpolation.

*Data and code availability*

Measurement data as well as the firmware and software code can be made available upon request.

**Acknowledgements**

The authors would like to thank Professor Ming Wu, the Marvell Nanofabrication Laboratory, Mohammad Meraj Ghanbari, the sponsors of the Berkeley Wireless Research Center, and the sponsors of the Berkeley Sensor & Actuator Center for the tools, equipment and helpful discussions. We are grateful to V. Aksyuk (NIST) for the useful discussions and to D. Czaplewski from Argonne National Laboratory for his help with the metallization process. This work was performed, in part, at the Center for Nanoscale Materials, a U.S. Department of Energy Office of Science User Facility and supported by the U.S. Department of Energy, Office of Science, under Contract No. DE-AC02-06CH11357.

Certain commercial equipment, instruments, or materials (suppliers or software) are identified in this paper to foster understanding. Such identification does not imply recommendation or endorsement by the National Institute of Standards and Technology, nor does it imply that the materials or equipment identified are necessarily the best available for the purpose.

**Conflict of interests**

The authors have no conflicts of interest to report.

**Contributions**

NTE designed the micromirror device, performed the characterization and testing, analysed the results, and wrote the manuscript. CY contributed to the concept development and built the firmware and software for the DAC-based voltage drive. NA was the main contributor to the design and assembly of the optical test setup. NP contributed to the concept development and provided the basis for the computational design framework. LW contributed to the concept development and provided resources for optical testing. DL contributed to the concept development, MEMS design, fabrication (including postprocessing), and array inspection under optical microscopy. RM contributed to the concept development, provided electronic resources for driving and characterization, supervised the execution of this work overall, and edited the manuscript. All the authors also provided manuscript feedback.